# Surface enhanced resonance Raman and luminescence on plasmon active nanostructured cavities

Frances Lordan, James H Rice
*School of Physics, University College Dublin, Belfield, Dublin 4, Ireland*

Bincy Jose, Robert J Forster, Tia E Keyes
*School of Chemical Sciences, Dublin City University, Dublin 9, Ireland*

## Abstract

Presented here are studies of the impact of excitation angle on surface enhanced Raman and luminescence spectroscopy of dye immobilised on a plasmon active nanocavity array support. Results show that both Raman and luminescence intensities depend on the angle of incidence consistent with the presence of cavity supported plasmon modes. Dependence of scattering or emission intensity with excitation angle occurs over the window of observation.

Presently there is significant interest in characterising and applying metallic nanostructured materials as optical imaging and sensing platforms.[1,2] The interaction of light with a metal can be controlled by creating nanoscale features on the metals surface. Such engineering of a metals surface can create localised or propagating surface plasmons. Nanofabrication methods can lead to architectures whereby the localised plasmons can be positioned on a metals surface with great precision and reproducibility and, most importantly for analytical application, can be addressed using far-field excitation.

Raman and fluorescence spectroscopies are well established analytical methods widely applied across chemical and biological substrates.[3,4] Surface enhanced Raman (SER) or luminescence enables significant increases in both the Raman and luminescence signal strength to be obtained.[5] However, the initial promise of SER and luminescence spectroscopies has largely remained undelivered due to issues regarding the reproducibility of the plasmonic metal platforms exploited for such methods. It is noted that while considerable developments have been made towards producing reproducible SERS platforms, such platforms applied to luminescence are less widely studied. Further improvement in understanding the way in which such substrates interact with adsorbates and incident and emitted light is needed so that their commercial (bio)analytical application potential can be fully realised.[6] In the case of nanocavity arrays, understanding the optimal region of plasmonic enhancement for luminescence and Raman spectroscopy and understanding how the angle of incident excitation affects enhancement are both of key importance.

We present here a study on a plasmon active array with reproducible nano-sized spherical cap cavity architectures on gold on fluorine doped tin oxide (FTO). Such nanocavity array substrates can be made cost-effectively and reproducibly using a wide range of materials.[6-8] The resulting voids or cavities have been shown to enable better energy confinement than metallic particle nanostructures, resulting in better surface enhancements.[7,8] The optical properties of these substrates can be readily controlled through sphere diameter and thickness of the electrodeposited film.[6,9] Metallic sphere segment void substrates sustain surface plasmons, they generate intense electric fields under illumination and consequently show huge enhancements of SERS in air and in contact with solution.[7,10]

To date, while studies have been performed in regard to the Raman properties of these substrates, few studies have been performed on luminescence enhancements by these substrates. Furthermore, whereas the angle dependence of surface enhanced Raman signal from such substrates have been widely studied by Bartelett, Cole and Baumberg et al. We present the the angle dependence of both SERRS and luminescent signal which were collected simultaneously. In addition we present resonance Raman studies which were recorded using these substrates. Previous Raman studies have concentrated on coupling the Raman scattering energy with a plasmon absorption frequency under non-resonance Raman scattering conditions.[7]

Gold nanocavity arrays on FTO substrate were prepared using a nanosphere lithographic technique based on electrodeposition through the voids of self-assembled polystyrene spheres on FTO as described previously.[10] The resulting gold nanocavities were investigated using SEM (Scanning Electron Microscopy) and AFM (Atomic Force Microscopy) measurements.

[Ru(bpy)$_2$(Qbpy)]$^{2+}$ was synthesized from cis-[Ru(bpy)$_2$Cl$_2$] as described previously.[11] The gold nanocavity array was sonicated in a 1 mM methanol solution of [Ru(bpy)$_2$(Qbpy)]$^{2+}$ for 30 minutes. Then the arrays were kept in dye solution for 2 days to form monolayers of dye molecules. Excess physisorbed material was then removed from the electrode by sonicating in methanol and rinsing prior to the measurements. The sample was mounted onto a customized goniometer at a tilt angle of 6 degrees. The sample was excited at 532 nm with the laser focused using a 10 cm focal length lens. The Raman and luminescence signals were collected at a backscattered angle and directed onto an EMCCD (Electron Multiplying Charge Coupled Device) via a monochromator. Spectra were accumulated for 20 seconds. The pump angle was varied and the SERS or luminescence was then measured.

The gold nanocavity samples studied here were made using 820 nm sized polystyrene spheres to form the templates. The surface features possess an aspect ratio (of depth of coverage to diameter) of $\bar{t}=0.76$ (see Fig 1(a)) using the following equation:[12]

$$\bar{t} = \frac{r_{void} \pm \sqrt{(r_{void}^2 - r_{pore}^2)}}{2 r_{void}} \tag{1}$$

Here, $r_{void}$ is the radius of the spherical cavity and $r_{pore}$ is the radius at the opening of the cavity. As reported previously,[10] the depth of the wells has been measured by cutting the sample and imaging the substrate side-on using a SEM. The measured depth of the cavity was determined in this way to be 600 nm. AFM was used to measure the width of the cavity which was determined to be 700 nm.

An AFM topography image recorded of the sample is shown in Fig 1(b). The image shows the presence of regular arrays of nanocavities on the sample surface. The AFM studies showed that the substrate possessed a rough surface with size features varying on average < 40 nm. This is a feature of the underlying FTO coated substrate.[13]

The Raman and luminescence spectra recorded using a 532 nm excitation wavelength are shown in Fig 1(c) and 1(d). The Raman spectra show several peaks which are characteristic of pyridine modes observed previously from surface enhanced Raman spectra of [Ru(bpy)$_2$(Qbpy)]$^{2+}$.[11] The luminescence spectrum was centered at c.a. 1.82 eV. This is in agreement with studies. [Ru(bpy)$_2$(Qbpy)]$^{2+}$ on planar gold and metallic cavities has been reported to emit in this region.[14]

The surface enhancement of the Raman spectra of [Ru(bpy)$_2$(Qbpy)]$^{2+}$ on the FTO coated gold array were reported previously to be approximately 10$^7$ under 514 nm excitation and comparable enhancement is anticipated here.[6] As described by Bartlett and Baumberg et al, Mie and Bragg plasmon distributions are influenced by excitation angle to different degrees.[7] Their studies of SER of benzenethiol were performed using silver nanocavites formed on a glass substrate. The cavities were 700 nm in diameter with a $\bar{t}=0.9$. SER studies were undertaken at 633 nm. The samples studied here were 820 nm cavities with $\bar{t}=0.76$. The excitation wavelength used here was 532 nm. This wavelength is post-resonant with the absorption band for $[Ru(bpy)_2(Qbpy)]^{2+}$ and consequently the Raman signal is anticipated to be both surface and resonantly enhanced. .

A series of surface enhanced resonance Raman (SERR) spectra were recorded as a function of incident angle. Fig 2(a) shows a plot of Raman intensity for the Raman band at 2.208 eV as a function of angle. Monitoring the change in the intensity of the Raman transition shows that the intensity of the SERR band depends upon angle. The intensity profile shows regions of relatively high Raman signal and regions of relatively low Raman signal which vary with θ. The intensity of the Raman scatter rises rapidly after θ = 0 degrees and then falls after θ = 20 degrees.

The luminescence intensity at 1.8 eV from the monolayer at the substrate was monitored as a function of θ. Fig 3 shows that the intensity of luminescence varies with positive and negative values of θ. his plot indicates that the luminescence observed is plasmon assisted. The maximum intensity of luminescence is seen at c.a. θ = +/-14 degrees.

The sample has cavities with dimensions ($\bar{t}$ = 0.76) that are expected to result in the formation of cavity localized (Mie) plasmons, localized (rim) plasmons on the top surface of the array and delocalized (Bragg) plasmons.[12,15] Each plasmon mode has a unique field distribution within the substrate. The intensities of the plasmons vary as a function of both θ and the frequency of incident EM radiation.[12,15] Both experimental and theoretical studies have indicated that the absorption energy of such cavity related plasmons change as a function of θ for angle of incidence.[8,12,15] Kelf et al simulated plasmonic intensities for 600 nm gold nanocavities as a function of angle (with respect to incident excitation) and cavity size.[12] The dependence of the energies of both Bragg and Mie plasmons demonstrated that the energy of these plasmon modes was sensitive to cavity size and cavity height. In addition strong interactions between the different plasmon modes as well as other mixing processes were reported.[12] The authors noted that different thickness regimes could be identified. At $\bar{t}$=0.4–0.9 Mie plasmons are present and mix with the Bragg modes. A previous study of angle dependence of SER using a nanocavity array of $\bar{t}$=0.5 by Bartlett et al used non-resonance excitation at $\lambda_{ex}$ = 632 nm. This excitation frequency was chosen for resonant plasmon enhancement of the SERs signal. Simulations of the plasmon dispersion for 600 nm cavites by Kelf et al[12] showed that at $\bar{t}$=0.5 strong plasmon bands were present around 1.8 eV i.e. the energy of Raman from $\lambda_{ex}$ = 632 nm. Our studies produce resonance Raman scattering energies of c.a. 2 eV. This indicates that the Raman is not in effective resonance with a plasmon band. In contrast to this the luminescence at 1.8 eV is anticipated to be resonant with the metal plasmon bands.

The Mie plasmons and rim plasmons are localized precisely on the substrate.[12,15] As a consequence of this the proximity of the molecule to each mode will determine whether SERS or luminescence will be observed. Each can be monitored separately at $\lambda_{ex}$ = 532 nm. It is known that molecules in direct contact with metallic surfaces exhibiting plasmons will result in SER while a separation of c.a. 10 nm is required between the molecule and the metallic surface for optimal luminescence. The nanocavity sample can simultaneously support localized plasmon modes for both SERS and luminescence. Nonetheless, as reported previously $[Ru(bpy)_2(Qbpy)]^{2+}$ exhibits significant emission when immobilized on gold. .[14,11] A further advantage of using [Ru(bpy)$_2$(Qbpy)]$^{2+}$ in these studies is that both Resonance Raman and emission can be excited at the same wavelength without interference from emission in the Raman signal due to the large Stokes shift of this dye. The nanocavity sample can simultaneously support localized plasmon modes for both SERS and luminescence.

In general, increases in SERR signal occur in concert with an increase with luminescence signal. This is explained via the mechanism for plasmon enhancement. Two different mechanisms are reported in the literature to explain the origin of these effects i.e. the charge transfer model and the electromagnetic model. The latter arises from the incident photon interacting with the substrate, thereby creating a plasmon. The presence of the plasmon in proximity to a molecule on the substrate surface causes the molecule to become polarized creating an effective dipole moment. Depending on the distance of the molecule from the plasmon this leads either to SERS or luminescence. Studies of these sphere segment void substrates have shown that they are capable of supporting SERS with enhancements of $>10^6$.[16]

Presented here are studies of incident angle dependence of SERR and luminescence intensity of a Ru complex on a plasmon active nanocavity support. Dependence of scattering or emission intensity on excitation angle is seen over the window of observation (c.a. 1.9 to 2.2 eV). Results show that luminescence enhancement possess angle dependence with this angle dependence confirming plasmonic enhancement. In addition studies here have shown that angle dependence for luminescence were observed on the nanocavity array despite the surface roughness of c.a. 40nm as observed using AFM.

The authors would like to acknowledge Science Foundation Ireland for supporting this research.

**References**

[1] Y. Lu, G. L. Liu, and L. P. Lee, Nano Lett. **5**, 5 (2005).

[2] M. Schnippering, H. V. Powell, M. Zhang, J. V. Macpherson, P. R. Unwin, M. Mazurenka, and S. R. Mackenzie, J. Phys. Chem. C, **112**, 15274 (2008).

[3] W. E. Smith, Chem. Soc. Rev. **5**, 955 (2008).

[4] R. S. Davidson, Chem. Soc. Rev. **4**, 241 (1996).

[5] G. Lajos, D. Jancura, P. Miskovsky, J. V. Garca-Ramos, and S. Sanchez-Cortes, J. Phys. Chem. C, **112**, 12974 (2008).

[6] A. Kaminska, O. Inya-Agha, R.J. Forster, and T.E. Keyes, Phys. Chem. Chem. Phys. **10**, 4172 (2008).

[7] J. J. Baumberg, T. A. Kelf, Y. S. S. Cintra, M. E. Abdelsalam, P. N. Bartlett, and A. E. Russell, Nano Lett. **5**, 2262 (2005).

[8] T. V. Teperik, F. J. García de Abajo, A. G. Borisov, M. Abdelsalam, P. N. Bartlett, Y. Sugawara, and J. J. Baumberg, Nature Photonics **2**, 299 (2008).

[9] E.J. Tull, P.N. Bartlett, and K.R. Ryan, Langmuir **23**, 7859 (2007).

[10] B. Jose, R. Steffen, U. Neugebauer, E. Sheridan, R. Marthi, R. J. Forster, and T. E. Keyes, Phys. Chem. Chem. Phys. **11**, 10923 (2009).

[11] R.J. Forster, Y. Pellegrin, D. Leane, J. L. Brennan, and T. E. Keyes, J. Phys. Chem. C **111**, 2063 (2007).

[12] T. A. Kelf, Y. Sugawara, R. M. Cole, J. J. Baumberg, M. E. Abdelsalam, S. Cintra, S. Mahajan, A. E. Russell, and P. N. Bartlett, Phys., Rev. B, **74**, 245415 (2006).

[13] L. Ottaviano, M. Kwoka, F. Bisti, P. Parisse, V. Grossi, S. Santucci, and J. Szuber, Thin Solid Films **517**, 6161 (2009).

[14] R.J. Forster, and T.E. Keyes, J. Phys., Chem., B **102**, 10004 (1998).

[15] R. M. Cole, S. Mahajan, P. N. Bartlett, and J. J. Baumberg, Optics Express **17**, 13298 (2009).

[16] S. Mahajan, R. M. Cole, B. F. Soares, S. H. Pelfrey, A. E. Russell, J. J. Baumberg, and P. N. Bartlett, J. Phys., Chem., C **113,** 9284 (2009).

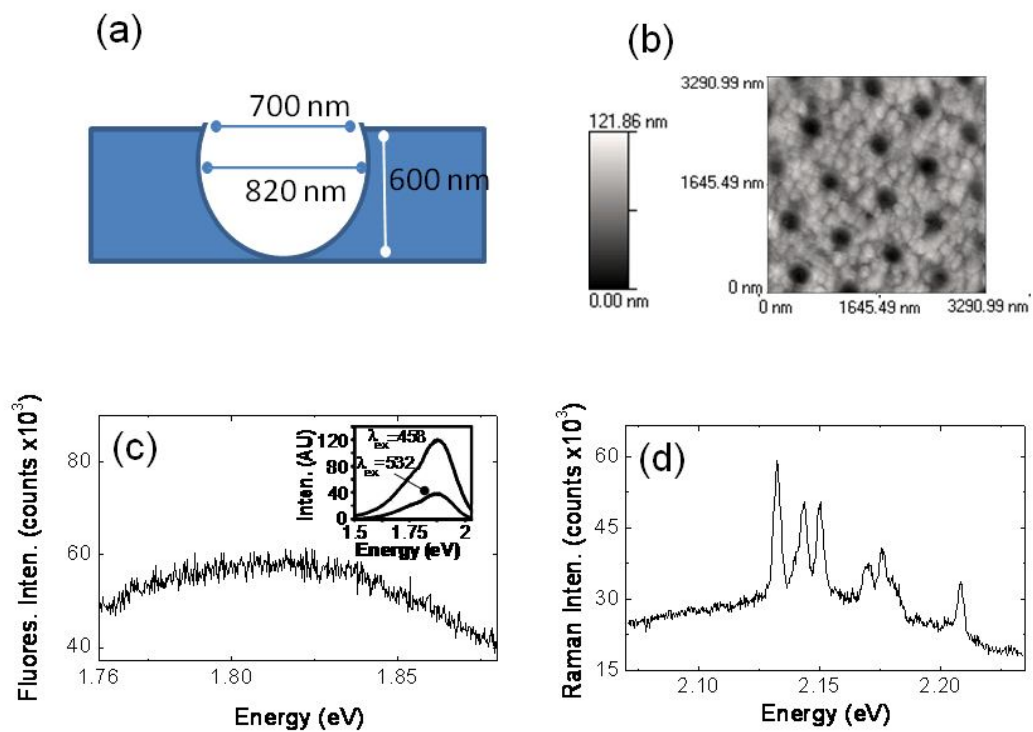

Fig 1. (a) schematic diagram of the cavity dimensions, (b) AFM image of the cavity showing cavity features as outlined in a), in addition it shows the presence of a regular array of cavities across the sample, (c) luminescence spectrum from the sample, insert, luminescence spectra from [Ru(bpy)2Qbpy)](ClO4)2 $1\times10^{-5}$M in ethanol/water (1/9 v/v) under 458 and 532 nm excitation, (d) typical Raman spectrum taken from the sample at +8 degrees. Excitation wavelength 532 nm, laser power 38 mW. Accumulation time 40sec.

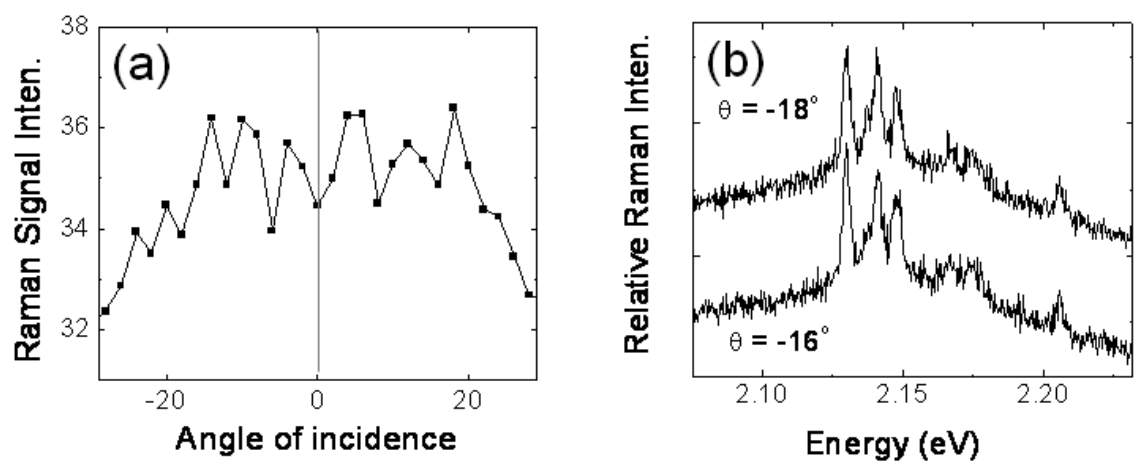

Fig 2. (a) a plot of Raman intensity as a function of angle for Raman band at 2.208 eV. (b) Raman spectra recorded at two angles of incidence, the Raman spectra are artificially off-set for clarity. Accumulation time 20sec.

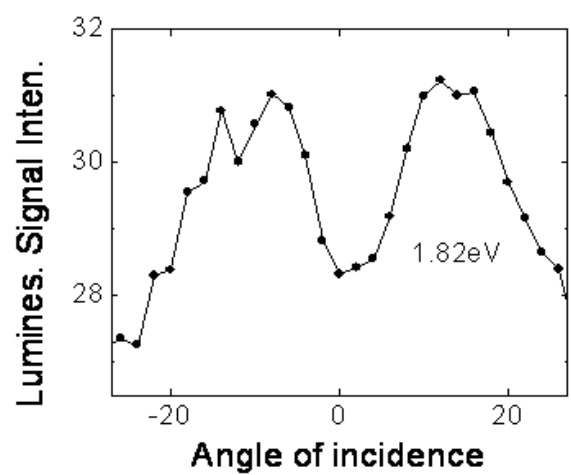

Fig 3. a plot of luminescence intensity as a function of angle.